\documentclass{PoS}

\PoS{PoS(LAT2005)163}

\title{The QCD phase diagram at finite density \\[-4cm]
\normalsize \tt \hspace*{11.3cm}BNL-NT-05/38\\
\normalsize \tt \hspace*{11.3cm}WUB-05-12\\
\normalsize \tt \hspace*{11.3cm}ITP-Budapest 623\\[2cm]}

\ShortTitle{The QCD phase diagram at finite density }

\author{\speaker{Christian Schmidt}\\
        Physics Department, Brookhaven National Laboratory, Upton, NY, 11973, USA\\
        E-mail: \email{cschmidt@bnl.gov}}

\author{Zoltan Fodor\\
        Department of Physics, University of Wuppertal, Wuppertal, Germany\\
        Institute for Theoretical Physics, E\"otv\"os University, Budapest, Hungary\\
        E-mail: \email{fodor@bodri.elte.hu}}

\author{Sandor Katz\\
        Institute for Theoretical Physics, E\"otv\"os University, Budapest, Hungary\\
        E-mail: \email{katz@bodri.elte.hu}}

\abstract{We study the density of states method to explore the phase diagram
of the chiral transition on the tempeature
and quark chemical potential plane. Four quark flavours are used in the
analysis. Though the method is quite expensive small lattices
show an indication for a triple-point connecting three different phases on
the phase diagram.}

\FullConference{XXIIIrd International Symposium on Lattice Field Theory\\
                 25-30 July 2005\\
                 Trinity College, Dublin, Ireland}

\def\lsim{\raise0.3ex\hbox{$<$\kern-0.75em\raise-1.1ex\hbox{$\sim$}}}
\def\gsim{\raise0.3ex\hbox{$>$\kern-0.75em\raise-1.1ex\hbox{$\sim$}}}

\begin{document}

\section{Introduction}
To clarify the phase diagram of QCD and thus the nature of matter under
extreme conditions is one of the most interesting and fundamental tasks of
high energy physics.
Lattice QCD has been shown to provide important and reliable information from
first principals on QCD at zero density. However, Lattice QCD at finite
densities has been harmed by the complex action problem ever since its
inception. For $\mu>0$ the determinant of the fermion
matrix ($\rm{det}M$) becomes complex. Standard Monte Carlo techniques using
importance sampling are thus no longer applicable when calculating observables
in the grand canonical ensemble according to the partition function
\begin{equation}
Z_{GC}(\mu)=\int \mathcal{D}U\; \rm{det}M[U](\mu) \exp\{-S_G[U]\}.
\label{eq:Z_GC}
\end{equation}
Recently many different methods have been developed to cirumvent the complex
action problem for small $\mu/T$ \cite{Fodor:2001au, methods}. For a recent overview see also \cite{overview}.

\section{Formulation of the method\label{sec:method}}
A very general formulation of the DOS
method is the following: One exposed parameter ($\phi$) is fixed. The
expectation value of a thermodynamic observable ($O$), according to the usual
grand canonical partition function (\ref{eq:Z_GC}), can be recovered by the
integral
\begin{equation}
<O>=\int d\phi \, \left<Of(U)\right>_\phi \rho(\phi)
\left/ \int d\phi \, \left<f(U)\right>_\phi \rho(\phi)\right.
\label{eq:dos_obs}
\end{equation}
where the density of states ($\rho$) is given by the constrained partition
function:
\begin{equation}
\rho(x)\equiv Z_\phi(x)=\int \mathcal{D}U\, g(U) \, \delta( \phi - x ).
\label{eq:dos_Z}
\end{equation}
With $\left<~\right>_\phi$ we denote the expectation value with respect to the
constrained partition function. In addition, the product of the weight
functions $f,g$ has to give the correct measure of $Z_{GC}$:
$fg=\rm{det}M\exp\{-S_G\}$. This idea of reordering the partition
functions is rather old and was used in many different cases
\cite{dos, LUO, Ambjorn}
The advantages of this additional integration becomes
clear, when choosing $\phi=P$ and $g(U)=1$. In this case $\rho(\phi)$ is
independent of all simulation parameters. The observable can be calculated as
a function of all values of the lattice coupling $\beta$. If one has stored
all eigenvalues of the fermion matrix for all configurations, the observable
can also be calculated as a function of quark mass ($m$) and number of
flavors\cite{LUO} ($N_f$). In this work we chose
\begin{equation}
\phi=P \qquad \mbox{and} \qquad
g=\left|{\rm det}M \right| \exp\{-S_G\}, \qquad f=\exp\{i\theta\}.
\label{eq:con}
\end{equation}
In other words we constrain the plaquette and perform simulations with measure
$g$. In practice, we replace the delta function in Equation~(\ref{eq:dos_Z}) by a
sharply peaked potential \cite{Ambjorn}. The constrained partition function
for fixed values of the plaquette expectation value can then be written as
\begin{equation}
\rho(x) \approx \int {\cal D}U\; g(U) \exp\left\{- V(x)\right\},
\end{equation}
where $\exp\{-V(x)\}$ is a Gaussian potential with
\begin{equation}
V(x)=\frac{1}{2}\gamma\left(x-P\right)^2.
\end{equation}
We obtain the density of states ($\rho(x)$) by the fluctuations of the actual
plaquette $P$ around the constraint value $x$. The fluctuation dissipation
theorem gives
\begin{equation}
\frac{d}{dx}\ln \rho(x)=<x-P>_x.
\end{equation}
Before performing
the integrals in Equation~(\ref{eq:dos_obs}) we compute from an ensemble generated at $(\mu_0,\beta_0)$:
\begin{eqnarray}
\label{eq:rew1}
\left<Of(U)\right>_x(\mu,\beta)
&=&\left<Of(U)R(\mu,\mu_0,\beta,\beta_0)\right>_x
/\left<R(\mu,\mu_0,\beta,\beta_0)\right>_x,\\
\label{eq:rew2}
\left<f(U)\right>_x(\mu,\beta)
&=&\left<f(U)R(\mu,\mu_0,\beta,\beta_0)\right>_x
/\left<R(\mu,\mu_0,\beta,\beta_0)\right>_x,\\
\label{eq:rew3}
\frac{d}{dx}\ln\rho(x,\mu,\beta)
&=&\left<(x-P)R(\mu,\mu_0,\beta,\beta_0)\right>_x.
\end{eqnarray}
Here $R$ is given by the quotient of the measure $g$
at the point $(\mu,\beta)$ and at the simulation point $(\mu_0,\beta_0)$,
\begin{equation}
R(\mu,\mu_0,\beta,\beta_0)=g(\mu,\beta)/g(\mu_0,\beta_0)
=\frac{|{\rm det}(\mu)|}{|{\rm det}(\mu_0)|}\exp\{S_G(\beta)-S_G(\beta_0)\}.
\end{equation}
Heaving calculated the expressions~(\ref{eq:rew1})-(\ref{eq:rew3}), we are
able to extrapolate the expectation value of the observable~(\ref{eq:dos_obs})
to any point $(\mu,\beta)$ in a small region around the simulation point
$(\mu_0,\beta_0)$. For any evaluation of $\left<O\right>(\mu,\beta)$, we
numerically perform the integrals in Equation~(\ref{eq:dos_obs}). We also
combine the data from several simulation points to interpolate between them.


\section{Simulations with constrained plaquette}
The value we want to constrain is the expectation value of the global
plaquette, which is given on every gauge configuration by the sum over all
lattice points ($y$) and directions ($\mu\nu$) of the local plaquette
$P_{\mu\nu}(y)$ and its adjoint $P^\dagger_{\mu\nu}(y)$,
\begin{equation}
P=\sum_y\sum_{1\le\mu<\nu\le4} \frac{1}{6}\left[
{\rm Tr}P_{\mu\nu}(y) + {\rm Tr}P^\dagger_{\mu\nu}(y) \right].
\end{equation}
Since the plaquette is also the main part of the gauge action,
\begin{equation}
S_G=-\beta\sum_x\sum_{1\le\mu<\nu\le4}\left\{\frac{1}{6}\left[
    {\rm Tr}P_{\mu\nu}(x) + {\rm Tr}P^\dagger_{\mu\nu}(x)\right]-1\right\} ,
\end{equation}
the additional potential $V$ can be easily introduced in the hybrid Monte
Carlo update procedure of the hybrid-R algorithm \cite{Gottlieb:mq}. 
After calculating 
the equation of motion for the link variables $U_\mu(y)$, 
we find for the
gauge part of the force
\begin{equation}
i\dot{H}_\mu(y)=\left[\frac{\beta}{3}U_\mu(y)T_\mu(y)
\left(1+\frac{\gamma(x-P)}{\beta}\right)\right]_{\rm TA}.
\label{force}
\end{equation}
Here the subscript ${\rm TA}$ indicates the traceless anti-Hermitian part of
the matrix.
We see that in each molecular dynamical step the
measurement of the plaquette is required. However, the only modification
in the gauge force is the factor in round brackets.

\section{The critical line and the determination of a triple-point\label{sec:line}}
Simulations have been performed with staggered fermions and $N_f=4$. We chose
9 differed points in the $(\beta,\mu)$-plane for the $4^4$ lattice and 8
points for the $6^4$ lattice. On each of these points we did simulations with
20-40 constrained plaquette values, all with quark mass $am=0.05$. Further
simulations has been done with $(\beta,\mu)=(5.1,0.3)$ on the $6^3\times8$ lattice
for $am=0.05$ and $am=0.03$.
  
\begin{figure}
\begin{center}
\begin{minipage}{.48\textwidth}
\raisebox{6.0cm}{(a)}\includegraphics[width=6.5cm, height=6.5cm]{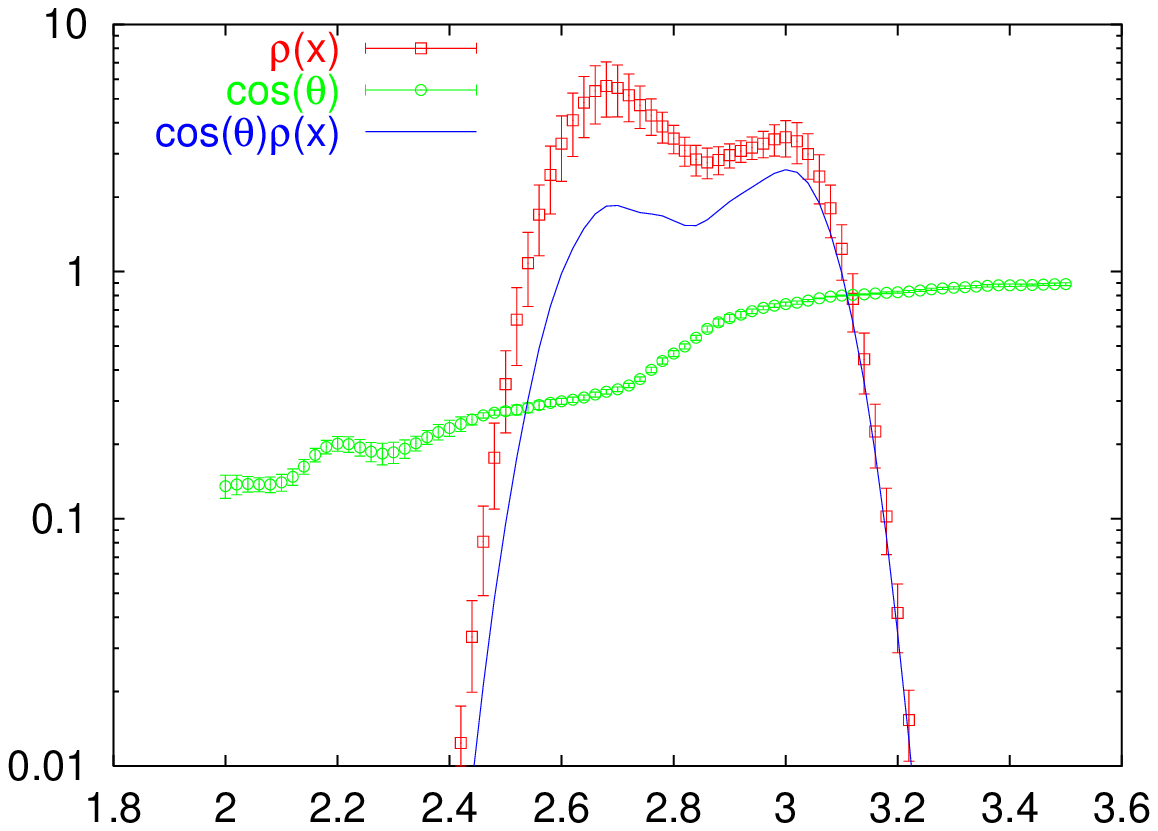}
\end{minipage}
\begin{minipage}{.48\textwidth}
\raisebox{6.0cm}{(b)}\includegraphics[width=6.5cm, height=6.5cm]{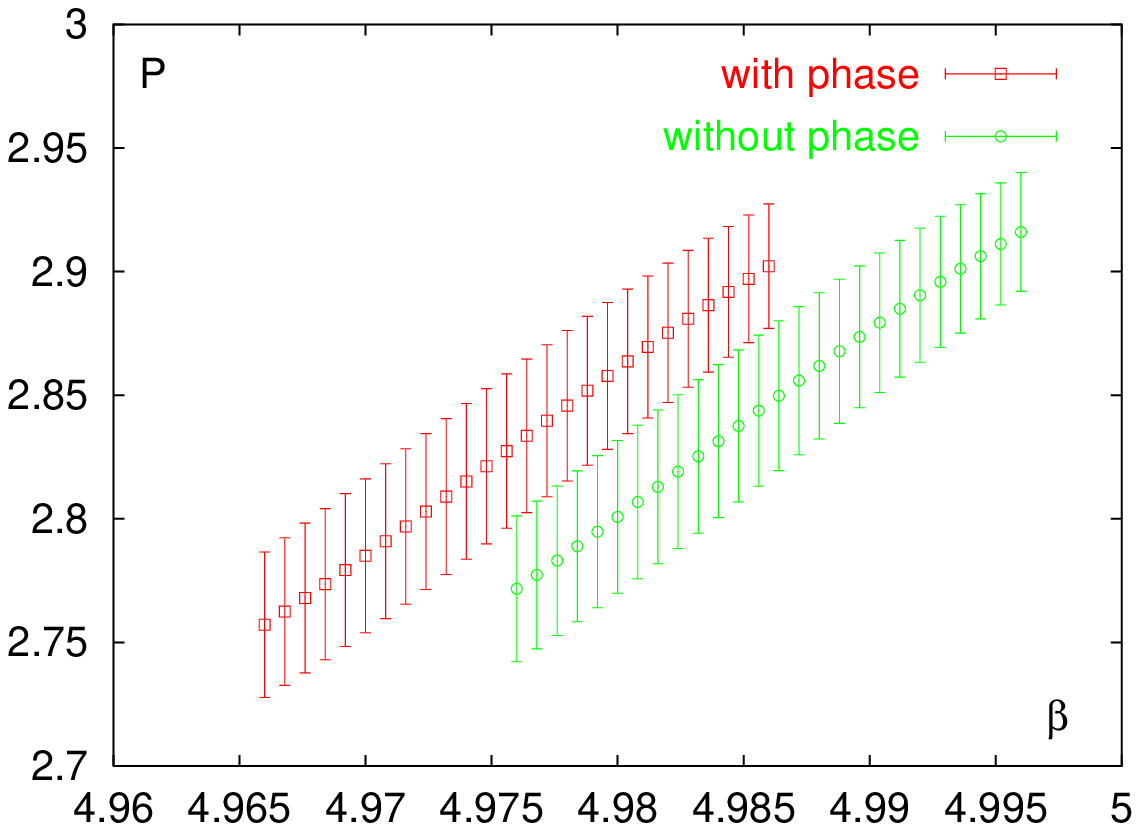}
\end{minipage}\\[4mm]
\begin{minipage}{.48\textwidth}
\raisebox{6.0cm}{(c)}\includegraphics[width=6.5cm, height=6.5cm]{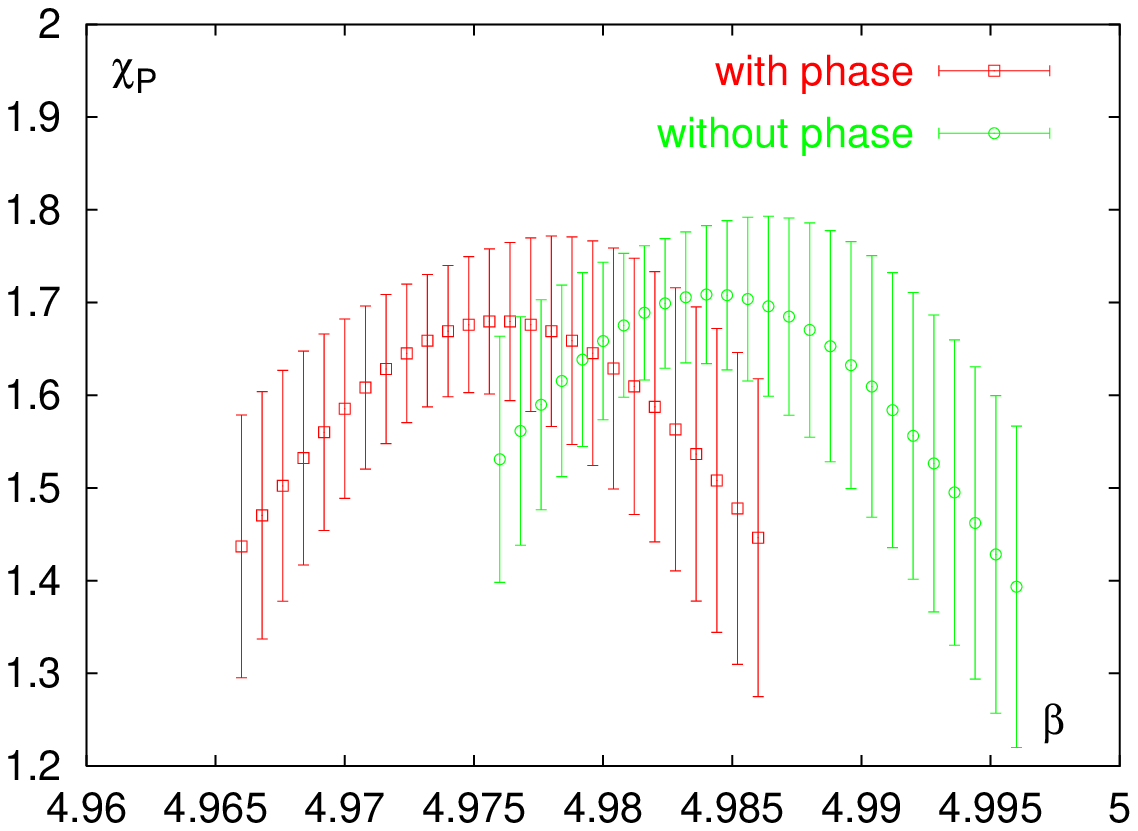}
\end{minipage}
\begin{minipage}{.48\textwidth}
\raisebox{6.0cm}{(d)}\includegraphics[width=6.5cm, height=6.5cm]{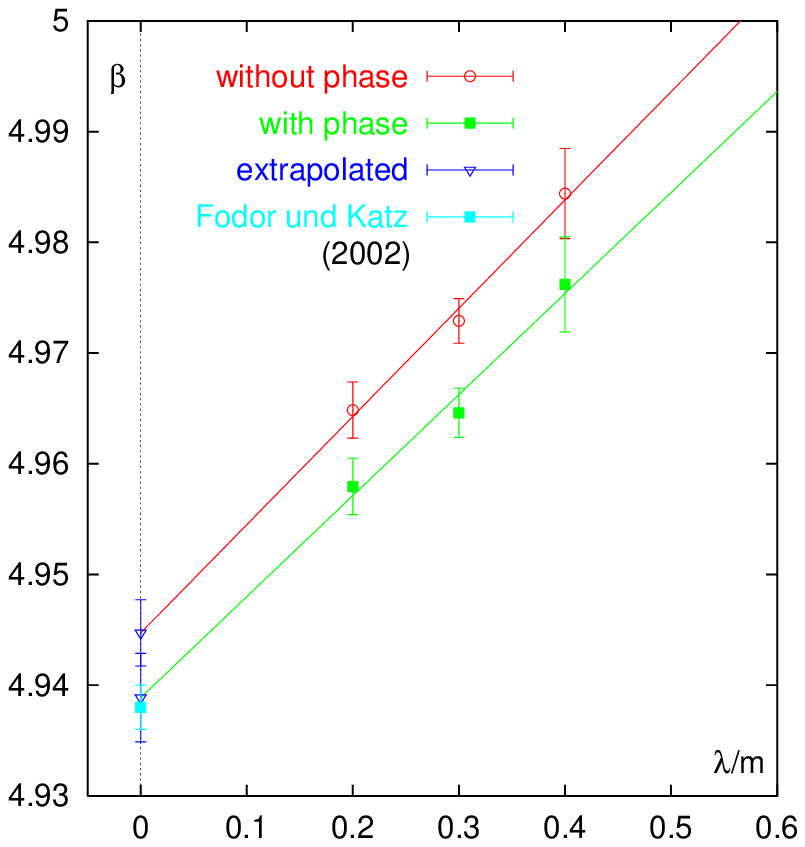}
\end{minipage}
\end{center}
\caption{Results for Simulations at $\beta=4.98$, $\mu=0.3$, $\lambda=0.02$,
  $n_f=4$, $am=0.05$, and number of lattice points: $4^4$. Shown are: (a) the density of
  states $\rho(x)$, the phase factor $\left<\cos(\theta)\right>$, and their
  product, (b) the Plaquette as a function of the coupling $\beta$, with and
  without the phase factor, (c) the Susceptibility of the Plaquette as a
  function of the coupling $\beta$, with and without the phase factor, and
  (d) the extrapolation $\beta_c(\lambda)$ to $\lambda=0$, with and without
  the phase factor.\label{fig:distri}} 
\end{figure}

Fist of all we check, whether we can reproduce old results with our new
method. We show in Figure~\ref{fig:distri}(a) results from a Simulation at 
$\mu=0.3$, $\beta=4.98$ and $\lambda=0.02$. 
Here we plot the density of states ($\rho$) and the real part of the phase
factor $\left<cos(\theta)\right>$ as a function of the constrained plaquette
value. The results have been interpolated in $P$, in order to obtain a better
result for the necessary integration over $P$. The distribution $\rho$ shows a
clear double peak structure, which signals the transition. The phase
factor is smaller in the low temperature phase ($P\lsim 2.8$). Hence in the
product $\rho\left<cos(\theta)\right>$ the low temperature peak is
suppressed. Now we perform the integrals
\begin{equation}
\left<P\right>=\int dx\; x \rho(x) \left<cos(\theta)\right>_x ,\qquad
\left<P^2\right>=\int dx\; x^2 \rho(x) \left<cos(\theta)\right>_x.
\end{equation}
In Figure~\ref{fig:distri}(b) we plot the plaquette expectation value $\left<P\right>$
as a function of the coupling $\beta$. 
The $\beta$-dependence is given by
Equations~(\ref{eq:rew1})-(\ref{eq:rew3}). We indeed find that including the 
phase factor does shift the transition to lower values of the coupling, which
also means to lower temperatures. This can also be seen in a shift of the peak
of the susceptibility of the plaquette
$\chi_P\equiv\left<P^2\right>-\left<P\right>^2$, which we plot in
Figure~\ref{fig:distri}(c). 
Since the $\lambda$ parameter introduces a systematic error, which can be
seen by the relative large critical coupling of $\beta_c=4.976(4)$ in
comparison to the result form multi-parameter reweighting $\beta_c=4.938(2)$
\cite{Fodor:2001au}, we perform the a linear extrapolation of $\lambda\to0$,
from $\lambda=0.02$, $\lambda=0.015$ and $\lambda=0.01$.
We show the extrapolation in Figure~\ref{fig:distri}(d). 
The extrapolated result $\beta=4.938(4)$ (including the phase factor) and the
result from multi-parameter reweighting are in very good agreement. 
From now on we only give results for $\lambda/m=0.2$, the $\lambda$ dependence is
however expected to be smaller for larger $\mu$. 

\begin{figure}
\begin{center}
\begin{minipage}{.48\textwidth}
\raisebox{6.0cm}{(a)}\includegraphics[width=6.5cm, height=6.5cm]{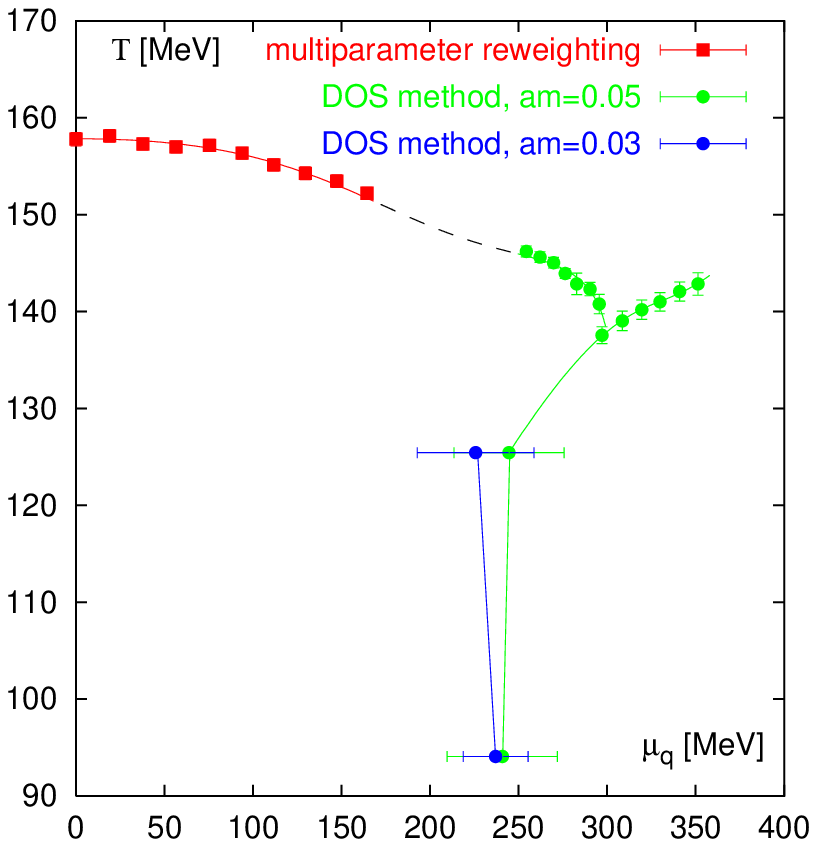}
\end{minipage}
\begin{minipage}{.48\textwidth}
\raisebox{6.0cm}{(b)}\includegraphics[width=6.5cm, height=6.5cm]{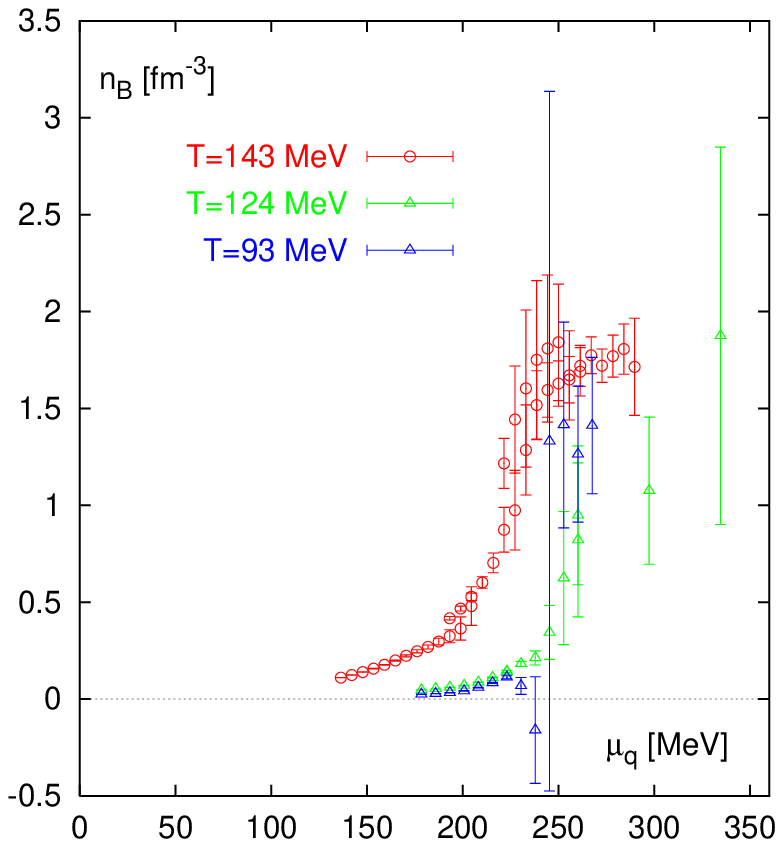}
\end{minipage}
\end{center}
\caption{The phase diagram in physical (a), and the quark number density at
  constant temperature $T=143~\mbox{MeV}$ 
  ($4^4$ lattice), $T=124~\mbox{MeV}$ ($6^4$ lattice) and $T=93~\mbox{MeV}$
  ($6^3\times 8$ lattice).
\label{fig:phase_diagram}}   
\end{figure}

In the range of $0.4\lsim a\mu\lsim0.5$ for the $4^4$ lattice, as well as $0.3\lsim
a\mu\lsim0.4$ for the $6^4$ lattice we found two transitions in the plaquette
expectation value $\left<P\right>(\beta)$. The two critical couplings result
in two transition lines in the phase diagram. The two transition lines are
almost perpendicular in the ($\beta, \mu$)-diagram, and join in a triple-point
of the phase diagram. in 
Figure~\ref{fig:phase_diagram}(a) we show the phase diagram in
physical units. 
The scale was set by the Sommer radius $r_0$, measured on a $10^3\times 20$
lattice. In both cases, the triple point is located around $\mu_q^{\rm tri}\approx
300~\mbox{MeV}$, however its temperature ($T^{\rm tri}$) decreases from $T^{\rm
  tri}\approx 148\mbox{MeV}$ on the $4^4$ lattice to $T^{\rm tri}\approx
137\mbox{MeV}$ on the $6^4$ lattice. 

Also shown in Figure~\ref{fig:phase_diagram}(a) are points from
simulations with quark mass $am=0.03$. The phase boundary turned out to be
--- within our statistical uncertainties --- independent of the the mass.
 
\section{The quark number density \label{sec:thermodynamics}}
To reveal the properties of the new phase located in the lower right corner of
the phase diagram, we calculated the quark number density, at constant coupling
$\beta$ and at constant temperature respectively. To obtain the density $n_q$
we perform the following integration
\begin{equation}
\label{eq:ddmu}
\left<\frac{{\rm d} \ln {\rm det}M}{{\rm d}(a\mu)}\right>
=\int dx\; \left<\frac{{\rm d} \ln {\rm det}M}{{\rm
      d}(a\mu)}cos(\theta)\right>_x \rho(x)
\label{eq:dmusq}
\end{equation}
The thermodynamic quantity $n_q$ are given as usual by
\begin{equation}
n_q = 
\frac{1}{a^3 N_s^3 N_t} 
\left<\frac{{\rm d} \ln {\rm det}M}{{\rm d}(a\mu)}\right>
\end{equation}
In Figure~\ref{fig:phase_diagram}(b) we show the baryon number density, which is related
to the quark number density by $n_B=n_q/3$.  
The results are plotted in
physical units and correspond to a constant temperature of $T\approx
143~\mbox{MeV}$ ($4^4$ lattice), $T\approx 124~\mbox{MeV}$ ($6^4$ lattice) and
$T\approx 93~\mbox{MeV}$ ($6^4\times 8$ lattice). In order to divide out the leading
order cut-off effect, 
we multiply we have multiplied the data with the factor $c=SB(N_t)/SB$, which
is the Stefan-Boltzmann value of a free lattice gas of quarks at a given value of $N_t$,
divided by its continuum Stefan-Boltzmann value. At the same value of the
chemical potential where we find also a peak in the susceptibility of the
plaquette ($\mu_c)$, we see a sudden rise in the baryon number density. Thus
for $\mu>\mu_c$ we enter a phase of dense matter. The transition occurs at a
density of $(2-3)\times n_N$, where $n_N$ denotes nuclear matter density. Above
the transition, the density reaches values of $(10-20)\times n_N$. Quite
similar results have been obtained recently by simulations in the 
canonical ensemble \cite{Alexandru:2004dx}.

\end{document}